\documentclass[floatfix,
twocolumn,
prb
]{revtex4}

\usepackage{graphicx}
\usepackage{rotating}
\usepackage{fancybox}

\usepackage[dvips]{pstcol} 

\begin{document}

\title{Visualizing proper-time in Special Relativity}

\author{Roberto B. Salgado}
\affiliation{Department of Physics, Syracuse University, Syracuse, New York 13244}
\email{salgado@physics.syr.edu}
\affiliation{Division of the Natural Sciences, Dillard University, New Orleans, Louisiana 70122}
\date{September 3, 2004}

\begin{abstract}

We present a new visualization of the proper-time elapsed along 
an observer's worldline. 
By supplementing worldlines with light 
clocks, the measurement of space-time intervals is reduced to the 
``counting of ticks.'' The resulting space-time diagrams are 
pedagogically attractive because they emphasize the relativistic 
view that ``time is what is measured by an observer's clock.'' 

\end{abstract}

\maketitle

\section{Introduction}

Einstein's special relativity
\cite{Ein1905}
forces us to revise our common-sense notions of time.
Indeed, clocks in relative motion will generally disagree on the elapsed
time interval measured between two meeting events.  The discrepancy is
practically undetectable for everyday relative speeds, but it is quite
significant when the relative speeds are comparable to the speed of light.
It is therefore necessary to distinguish these time intervals.
Thus, we define for each observer his ``proper-time'' 
as the elapsed time interval measured by his clock.
The goal of this paper is to find a physically intuitive
visualization of this concept of proper-time.

Many textbooks
\cite{FeynmanLectures,PerpendicularMirror,TaylorWheeler,MerminSTSR,ParallelMirror}
introduce proper-time
by analyzing the propagation of a light 
in a light clock,
\cite{Ein1905, vonLaue, Marzke, Operational, Arzelies, Arons}
which consists of a pair of mirrors
that face each other and are separated by a proper distance $L$.
One ``tick'' of this clock is the duration of one round trip
of a light ray bouncing back and forth between these mirrors.
The analysis is usually done in the context of
a simplified Michelson-Morley apparatus
\cite{Michelson}
whose arms may be regarded as light clocks.
Unfortunately, most of these presentations
\cite{FeynmanLectures, PerpendicularMirror, MerminSTSR} %
work in moving frames of
reference without making the connection to the 
space-time formulation,
first introduced by Minkowski%
\cite{Minkowski}
in 1907 %
and later extended by Einstein.%
\cite{Einstein-Spacetime}

Let us recall a quote from J.L.~Synge:
\begin{quotation}
        ...We have in the special theory of relativity the Minkowskian
        geometry of a flat 4-space with indefinite metric...
        Unfortunately, it has been customary to avoid this geometry,
        and to reason in terms of moving frames of reference, each with
        its own Euclidean geometry. As a result, intuition about Minkowskian
        space-time is weak and sometimes faulty....
	\cite{SyngeQuote}
\end{quotation}

Indeed, when studying observers in relative motion, 
it is advantageous to draw a spacetime-diagram
of the situation.
However, we are immediately faced with an important 
question: 
``how does one know where to mark off the 
ticks of each clock?''
More precisely, ``given a standard of time marked on an 
observer's worldline, how does one calibrate the same
standard on the other observer's worldline?''

\begin{figure}
\begin{center}
\begin{picture}(200,200)(0,0)
	\put(75,57){$O$}
	\put(96,178){\vector(0,1){10}}
	\put(100,185){$t$ (sec)}
	\put(155,54){$x$ (light-sec)}
	\resizebox{!}{3in}{
		\includegraphics[2in,3.2in][7in,8.4in]{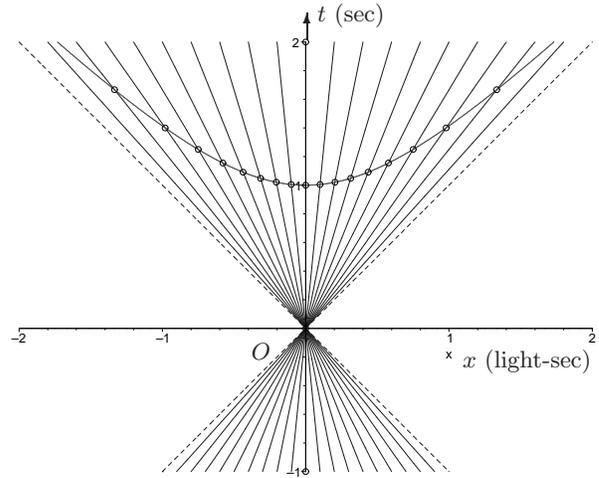}
	}
\end{picture}
\hspace*{-207pt}
\begin{picture}(200,150)(0,0)
\end{picture}
\vspace*{-0.25in}
\caption{\small For inertial observers meeting at event~$O$, 
the corresponding ticks of their clocks [synchronized at event~$O$]
trace out hyperbolas centered about event~$O$.
}
\label{fig:hyperbolaTicks}
\end{center}
\end{figure}

One approach is to use the invariance of the speed of light
to algebraically demonstrate 
the invariance of the spacetime-interval, 
from which the equation of a hyperbola arises.
\cite{TaylorWheeler, Minkowski}
Then, for inertial observers that meet
at a common event~$O$, it can be shown that
the corresponding ticks 
on their clocks [synchronized at event~$O$]
trace out hyperbolas on a spacetime diagram.
(See Figure~\ref{fig:hyperbolaTicks}.)
Once this result is established, 
many of the results of special relativity follow.  
This approach, however, is probably too sophisticated
for a novice.  Its connection with the 
standard textbook approach, using the
more familiar (though provisional)
physical concepts of time and space,
is not
readily apparent.

                      %

In this paper, we connect the two approaches
by drawing the spacetime-diagram of the 
Michelson-Morley apparatus.
Surprisingly, the only other 
spacetime-diagram of the apparatus
is a rough sketch in Synge's
{\em Relativity:~The~Special~Theory.}%
\cite{SyngeMM}

The resulting diagram provides 
a visualization of proper-time which 
explicitly incorporates the principle of relativity and
the invariance of the speed of light.
The standard ``effects'' of
time-dilation, length-contraction, and the relativity
of simultaneity are easily inferred from the diagram.
In addition, we show that standard calculations
\cite{FeynmanLectures, PerpendicularMirror, ParallelMirror, Schild, Rindler, Ellis}
of the Clock Effect and the Doppler Effect can be
reduced to the ``counting of ticks.''
We feel that the resulting diagrams are pedagogically attractive
since they emphasize the relativistic view that
``time is measured by an observer's clock.''

In the last section, we will consider a simplified version of our clock,
called the ``longitudinal light clock.''
Although this encodes fewer features than the full light clock, 
the longitudinal light clock is easy to draw manually.

In this paper, we have provided the detailed calculations 
used to draw the diagrams. However, we believe that
one can first qualitatively construct the diagram 
for the novice, emphasizing the physical principles first.  
Then, for those interested, one can continue quantitatively
with the analytical construction.

Following the standard conventions for spacetime diagrams,
time runs upward on our spacetime diagrams.
The scales of the axes are chosen so that
light rays are drawn at $45$ degrees.


\section{A simplified Michelson-Morley apparatus}

\subsection{An apparatus at rest}

Our simplified Michelson-Morley apparatus
has a light source at the origin and two mirrors,
each located a distance $L$ along {\sl a set of perpendicular arms}.

First, let us draw the spacetime diagram of the apparatus in its
inertial rest-frame, called the ``A-frame.''
The coordinates $(x,y,t)$ will be used to describe the events
from this frame.
Since relative motion will be taken to be along the $x$-axis,
the mirror along the $x$-axis will be called the ``longitudinal mirror''
and
the mirror along the $y$-axis will be called the ``transverse mirror.''

The worldlines of A's light source and mirrors
are described parametrically~by
\begin{eqnarray}
	\left.
   \begin{array}{ll}
     \mbox{A's light source}&
     \left\{
     \begin{array}{l}
       x(t)=0\\
       y(t)=0
     \end{array}
     \right.\\
\\
     \mbox{A's transverse mirror}&
     \left\{
     \begin{array}{l}
       x(t)=0\\
       y(t)=L
     \end{array}
     \right.\\

\\
     \mbox{A's longitudinal mirror}&
     \left\{
     \begin{array}{l}
    x(t)=L\\
    y(t)=0.
     \end{array}
     \right.\\
   \end{array}
   \right\}\label{eq:A-rest-worldlines}
\end{eqnarray}

Special relativity tells us that, in {\em all} inertial frames,
light travels through the vacuum with speed $c$ in all
spatial directions,
where
the speed $c$ has the value%
\cite{fn:speedOfLight}
of
$2.99792458\times 10^{8}\rm\ m/s$.

\begin{figure}
    \begin{center}
    \begin{picture}(200,270)(-30,0)
    \put(180,40){$x$}\put(2,220){$t$}\put(-40,57){$y$}
    \put(3,165){$T_A$}
    \put(-28,110){$Y_A$}
    \put(60,110){$X_A$}
    \put(73,209){\vector(-1,0){85}}
    \put(75,200){\shortstack{\small worldline of the\\\small transverse mirror}}
    \put(0,0){\resizebox{!}{3in}{
        \includegraphics{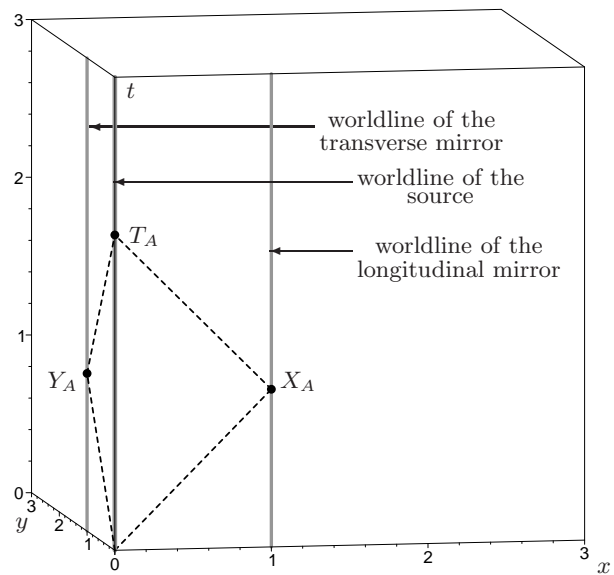}     
    	}
    }
    \end{picture}
    \hspace*{-207pt}
    \begin{picture}(200,270)(-30,0)
    \put(88,188){\vector(-1,0){90}}
    \put(90,180){\shortstack{\small worldline of the\\\small source}}
    \put(88,162){\vector(-1,0){31}}
    \put(90,152){\shortstack{\small worldline of the\\\small longitudinal mirror}}
    \end{picture}
    \vspace*{-0.5in}

    \caption{\small This is the spacetime diagram of a simplified
    Michelson-Morley apparatus in its rest-frame.
    (The $x$- and $y$-axes are marked in units of $L$. The $t$-axis is
    marked in units of $L/c$.)}
    \label{fig:rest}
    \end{center}
\end{figure}

Let event~$O$, with coordinates $(0,0,0)$, mark
the emission of a flash of light from the source.
One light ray emitted at event~$O$ reaches the
transverse mirror
at event
$$Y_A:\qquad \left( 0,\quad L,\quad \frac{L}{c} \right)$$
since light travels with speed $c$ for a time $L/c$ in order
to reach the transverse mirror a distance $L$ away.
Its reflection
is received back at the source
at event
$$T_A:\qquad \left(0 ,\quad 0,\quad 2\frac{L}{c} \right).$$
Similarly, one light ray
reaches
the longitudinal mirror
at event
$$X_A:\qquad \left(L,\quad 0,\quad \frac{L}{c} \right),$$
and its reflection is also received at event $T_A$.
Hence, the two rays, emitted at event~$O$ and
directed in different directions,
are received at a common event~$T_A$,
whose coordinates are $(0,0,2L/c)$.
(See Figure~\ref{fig:rest}.)

                      %

If one arranges another light ray to be emitted upon reception
(for example, by placing suitably oriented mirrors at the source),
then this apparatus can serve as a simple clock---the
light clock. %

The reception event $T_A$ marks one ``tick'' of this clock.
This tick will be chosen to be the ``standard tick.''
The elapsed time logged by an observer sitting at the source
is equal to the number of ticks multiplied by $2L/c$, the duration
of one round trip of a light ray.%
\cite{fn:ignoreMomentum}
If a finer scale of time is required,
one can increase the resolution by choosing a smaller separation~$L$.
In addition, the reflection events $X_A$ and $Y_A$,
which are on mirrors equidistant from the source,
can be regarded as ``half-ticks'' of this clock.
We define these half-ticks to be ``simultaneous events for this clock.''

%
%



\subsection{An apparatus in motion}

Now, suppose an identical apparatus
moves with spatial-velocity $v$ parallel to the $x$-axis of the A-frame.
This moving inertial-frame will be called the ``B-frame,''
and the coordinates~$(x',y',t')$ will be used to describe events from this frame.
For simplicity, the origins of the primed and unprimed
coordinate systems are taken to coincide at the emission event~$O$.
In addition, the corresponding spatial axes
are assumed to be spatially-parallel%
\cite{fn:aligningAxes}
within each inertial-frame.

Since B's apparatus is identical to A's,
the worldlines of B's light source and mirrors are described parametrically~by
\begin{eqnarray}
    \left.
   \begin{array}{ll}
     \mbox{B's light source}&
     \left\{
     \begin{array}{l}
       x'(t')=0\\
       y'(t')=0
     \end{array}
     \right.\\
\\
     \mbox{B's transverse mirror}&
     \left\{
     \begin{array}{l}
       x'(t')=0\\
       y'(t')=L
     \end{array}
     \right.\\
\\
     \mbox{B's longitudinal mirror}&
     \left\{
     \begin{array}{l}
    x'(t')=L\\
    y'(t')=0.
     \end{array}
     \right.\\
   \end{array}
   \right\}\label{eq:B-worldlines}
\end{eqnarray}

What does the spacetime diagram of this moving apparatus
look like in the A-frame?
In particular, what are the $(x,y,t)$ coordinates of the
worldlines of the moving apparatus and of the events~$X_B$,
$Y_B$, and $T_B$?

\begin{figure}[!ht]
    \begin{picture}(200,210)(0,0)
    \put(7,38){$O$}
    \put(210,32){$x$}
    \put(-3,173){$y$}
    \put(10,73){$L$}\put(136,73){$L$}
    \put(40,65){$ct_Y$}\put(100,65){$ct_Y$}
    \put(30,90){$vt_Y$}
    \put(110,90){$vt_Y$}

    \put(70,103){$Y_B$}
    \put(148,51){$T_B$}

    \put(140,32){$vt_B$}


    \resizebox{3in}{!}{
        \includegraphics*[59pt, 58pt][547pt, 443pt]{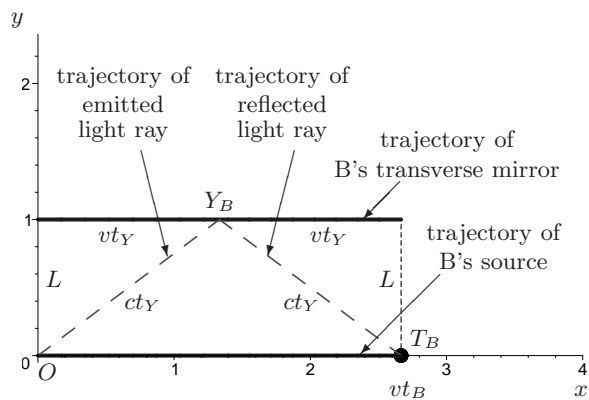}     
    }
    \end{picture}
    \hspace*{-207pt}
    \begin{picture}(200,210)(0,0)

    \put(155,80){\shortstack{\small trajectory of \\ \small B's source}}
    \put(160,78){\vector(-1,-1){30}}

    \put(120,115){\shortstack{\small trajectory of \\ \small B's transverse mirror}}
    \put(145,113){\vector(-1,-1){14}}

    \put(15,130){\shortstack{\small trajectory of \\ \small emitted \\ \small light ray}}
    \put(47,125){\vector(1,-4){10}}

    \put(75,130){\shortstack{\small trajectory of \\ \small reflected \\ \small light ray}}
    \put(105,125){\vector(-1,-4){10}}
    \end{picture}
    \vspace*{-0.5in}

    \caption{\small On A's $xy$-plane,
    the spatial trajectories of B's transverse arm and its associated light rays
    are drawn.
    The marked dot corresponds to the spatial coordinates of B's first ``tick,''
    which occurs after an elapsed time $t_B$ in the A-frame.
    Let $t_Y$ be the elapsed time for a light ray from event $O$ to reach
    B's transverse mirror.
    Using the Pythagorean theorem, it can be shown that $t_Y=(L/c)(1-(v/c)^2)^{-1/2}$.
    By symmetry, it follows that $t_B=(2L/c)(1-(v/c)^2)^{-1/2}$, which is
    longer than the duration of A's tick, $(2L/c)$.
    This is called the ``time dilation'' effect.
    (The $x$- and $y$-axes are marked in units of~$L$.) }
    \label{fig:XY}
    \end{figure}

{

Since B's light source and transverse mirror move with velocity $v$
in the $x$-direction,
they are described as:\cite{fn:noTransverseContraction}
\begin{eqnarray*}
   \begin{array}{ll}
     \mbox{B's light source}&
     \left\{
     \begin{array}{l}
       x(t)=vt\\
       y(t)=0.
     \end{array}
     \right.\\
%
\\
     \mbox{B's transverse mirror}&
     \left\{
     \begin{array}{l}
       x(t)=vt\\
       y(t)=L.
     \end{array}
     \right.\\
     \end{array}
\end{eqnarray*}
Due to this mirror's motion,
the light ray from event $O$ that meets this mirror
must travel a longer distance in the A-frame. 
(See Figure~\ref{fig:XY}.)
Using the Pythagorean theorem,
we find this distance is 
$L(1-(v/c)^2)^{-1/2}$,
which is
traveled by light in time 
$(L/c)(1-(v/c)^2)^{-1/2}$.
Thus, the reflection event on B's transverse mirror
is
$$Y_B:\quad \left(  v \frac{L}{c} \frac{1}{\sqrt{1-(v/c)^2}} , \quad L ,\quad \frac{L}{c}\frac{1}{\sqrt{1-(v/c)^2}} \right).$$
Similarly,
the reflected ray is received by B's source at:
$$T_B:\quad \left(  v \frac{2L}{c} \frac{1}{\sqrt{1-(v/c)^2}} , \quad 0,\quad \frac{2L}{c}\frac{1}{\sqrt{1-(v/c)^2}} \right).$$
Note that, in the A-frame, event~$T_B$
(``the first tick of the moving clock'') occurs later than
$T_A$
(``the first tick of the stationary clock'').
This is the ``time dilation'' effect.

}

                      %
%

\enlargethispage*{15\baselineskip}
\smallskip

\begin{figure}[!b]
    \begin{center}
    \begin{picture}(200,270)(-35,0)
    \put(-32,39){$O$}
    \put(170,30){$x$}
    \put(-45,260){$t$}
    \put(-55,221){$t_B$}
    \put(-55,204){$t_X$}
    \put(-55,152){$t_A$}
    \put(32,214){$vt_B$}
    \put(22,197){$vt_X$}        
    \put(102,196){$\ell$}       \put(-7,30){$\ell$}  \put(17,30){$L$}
    \put(110,245){$c(t_B-t_X)$}
    \put(127,200){$X_B$}
    \put(96,229){$T_B$}
    \put(118,239){\vector(-1,-4){4}}

    \put(120,30){$ct_X$}

    \put(28,60){\shortstack{\small worldline of \\ \small B's longitudinal mirror}}


    \put(0,0){\resizebox{!}{3in}{
        \includegraphics{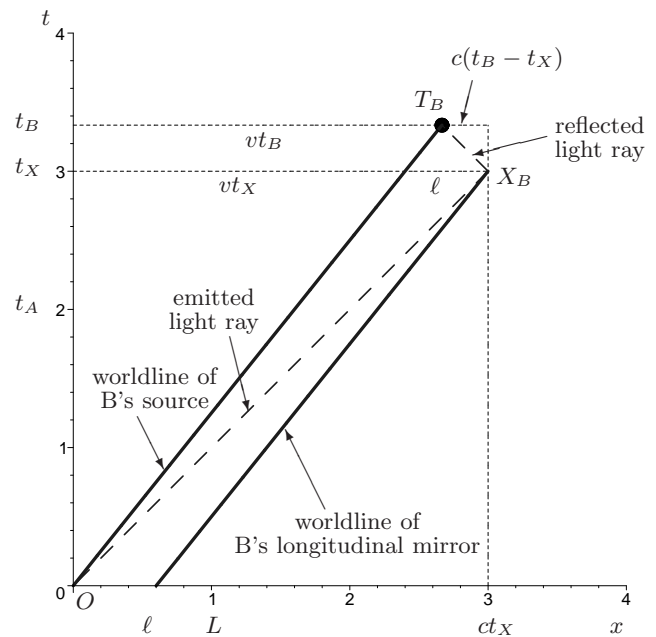}   
   		}
    }
    \end{picture}
    \hspace*{-207pt}
    \begin{picture}(200,270)(-35,0)

    \put(4,145){\shortstack{\small emitted \\ \small light ray}}
    \put(27,142){\vector(1,-4){7}}

    \put(150,210){\shortstack{\small reflected \\ \small light ray}}
    \put(147,218){\vector(-4,-1){28}}

    \put(63,77){\vector(-1,2){15}}

    \put(-25,115){\shortstack{\small worldline of \\ \small B's source}}
    \put(-8,112){\vector(1,-2){10}}

    \end{picture}
    \vspace*{-0.5in}
    \caption{\small On A's $xt$-plane,
    the worldlines of B's longitudinal arm and its associated light rays
    are drawn.
    The marked dot corresponds to B's first ``tick.''
    Let $\ell$ be the apparent length of B's longitudinal arm.
    Let $t_X$ be the elapsed time for a light ray from event $O$ to reach
    B's longitudinal mirror.
    Since $ct_X=vt_B+c(t_B-t_X)$ and $t_B=(2L/c)(1-(v/c)^2)^{-1/2}$,
    it can be shown that $t_X=(L/c)((1+v/c)/(1-v/c))^{1/2}$.
    Furthermore, since $ct_X=vt_X+\ell$, it follows that $\ell=L(1-(v/c)^2)^{1/2}$, which is
        shorter than the proper length $L$ of A's identical apparatus.
    This is the ``length contraction'' effect.
    (The $x$-axis is marked in units of $L$. The $t$-axis is
    marked in units of $L/c$.)}
    \label{fig:XT}
    \end{center}
\end{figure}

Now, let us consider the longitudinal mirror. (See Figure~\ref{fig:XT}.) 
Recall that the reflected light rays
for A's apparatus were received by A's source
at a common event $T_A$.
According to the principle of relativity,
the apparatuses cannot distinguish their states of inertial motion.
Thus, the reflected light rays
for B's apparatus must also be received by B's source
at a common event, here,~$T_B$.

                      %

\clearpage

For light rays to be received at event~$T_B$, what is the
required reflection event $X_B$ on B's longitudinal mirror?
Event~$X_B$ is the intersection on the $xt$-plane
of the forward-directed light ray from event~$O$ and
the backward-directed light ray toward event~$T_B$.
(Refer again to Figure~\ref{fig:XT}.)
After a little algebra, the reflection event $X_B$
is determined to be
$$X_B:\quad \left( c\frac{L\sqrt{1-(v/c)^2}}{c-v} ,
\quad 0, \quad \frac{L\sqrt{1-(v/c)^2}}{c-v} \right).$$
Thus, B's longitudinal mirror is described by:
\begin{eqnarray*}
   \begin{array}{ll}
     \mbox{B's longitudinal mirror}&
     \left\{
     \begin{array}{l}
       x(t)=vt+L\sqrt{1-(v/c)^2}\\
       y(t)=0.
     \end{array}
     \right.\\
     \end{array}
\end{eqnarray*}
Note that, in the A-frame,
the length of B's longitudinal arm is $L(1-(v/c)^2)^{1/2}$,
which is shorter than its proper length $L$.
This is the ``length contraction'' effect.

This completes the construction of B's apparatus.


As a check, these results can be obtained directly from the Lorentz transformation:
\begin{eqnarray}
    \left.
   \begin{array}{ll}
   t' &=\displaystyle \frac{ t-vx/c^2 }{\sqrt{1-(v/c)^2}}\\
   x' &=\displaystyle \frac{ x-vt }{\sqrt{1-(v/c)^2}}\\
   y' &=\displaystyle \vphantom{\frac{y}{y}} y
   \end{array}
   \right\}\label{eq:LT}
\end{eqnarray}
For instance, given the worldlines for B's apparatus
in $(x',y',t')$-coordinates (Equation~\ref{eq:B-worldlines}),
expressions for $x$ and $y$ as functions of $t$ can be obtained:
\begin{eqnarray}
	\left.
   \begin{array}{ll}
     \mbox{B's light source}&%
\hspace*{-0.08in}%
\left\{
     \begin{array}{l}
       x(t)=vt\\
       y(t)=0
     \end{array}
     \right.\\
     \mbox{B's transverse mirror}&%
\hspace*{-0.08in}%
\left\{
     \begin{array}{l}
       x(t)=vt\\
       y(t)=L
     \end{array}
     \right.\\
     \mbox{B's longitudinal mirror}&%
\hspace*{-0.08in}%
\left\{
     \begin{array}{l}
    x(t)=vt+L\sqrt{1-(v/c)^2}\\
    y(t)=0.
     \end{array}
     \right.\\
   \end{array}%
\hspace*{-0.1in}%
\right\}\label{eq:B-rest-worldlines}
\end{eqnarray}
Similarly, given the $(x',y',t')$-coordinates of B's tick and half-ticks,
the $(x,y,t)$-coordinates of $X_B$, $Y_B$, and $T_B$ can be obtained.



\begin{figure}
\begin{center}
\begin{picture}(200,280)(-30,0)
	\put(180,40){$x$}\put(2,220){$t$}\put(-40,57){$y$}
	\put(0,110){$T_A$}
	\put(-24,84){$Y_A$}
	\put(55,70){$X_A$}
	\put(60,156){$T_B$}
	\put(114,131){$X_B$}
	\put(35,140){$Y_B$}
	\put(40,135){\vector(-1,-4){7}}
	\put(130,160){\vector(-1,0){17}}
	\put(132,150){\shortstack{\small not length \\ \small contracted}}
	\put(-11,234){\small ``stationary''}
	\put(120,238){\small ``moving''}
	\resizebox{!}{3in}{
		\includegraphics{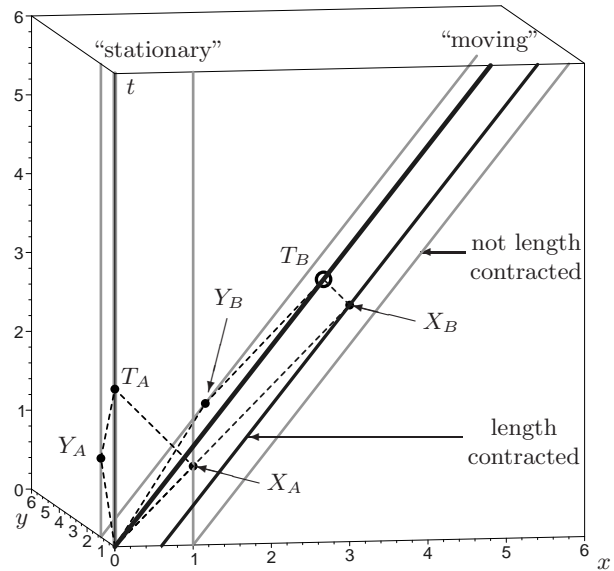}   
	}
\end{picture}
\hspace*{-207pt}
\begin{picture}(200,200)(-30,0)
	\put(130,90){\vector(-1,0){81}}
	\put(132,80){\shortstack{\small length \\ \small contracted}}
	\put(53,73){\vector(-4,1){24}}
	\put(112,134){\vector(-4,1){24}}
\end{picture}
\vspace*{-0.5in}
\caption{\small
This is A's spacetime diagram of B's
identical apparatus, which moves with velocity $v=0.8c$ along A's $x$-axis.
}
\label{fig:contract}
\end{center}
\end{figure}

These results are summarized in Figure~\ref{fig:contract}.

                      %

\enlargethispage*{5\baselineskip}

In addition to the time-dilation and length-contraction effects,
note that the events~$X_B$ and~$Y_B$, 
which are defined to be simultaneous according to B's clock,
are not simultaneous according to A's clock.
This is the ``relativity of simultaneity.''

{
\samepage

                      %
\begin{figure}
\begin{center}
    \begin{picture}(200,280)(-30,0)
    	\put(180,40){$x$}\put(2,220){$t$}\put(-40,57){$y$}
    	\put(0,110){$T_A$}
    	\put(-24,84){$Y_A$}
    	\put(55,70){$X_A$}
    	\put(144,190){$X_{Gal}$}
    	\put(60,156){$T_B$}
    	\put(102,220){$U_{Gal}$}
    	\put(35,140){$Y_B$}
    	\put(40,135){\vector(-1,-4){7}}
    	\put(130,160){\vector(-1,0){17}}
    	\put(132,150){\shortstack{\small not length \\ \small contracted}}
    	\put(-11,234){\small ``stationary''}
    	\put(120,238){\small ``moving''}
    	\resizebox{!}{3in}{
    		\includegraphics{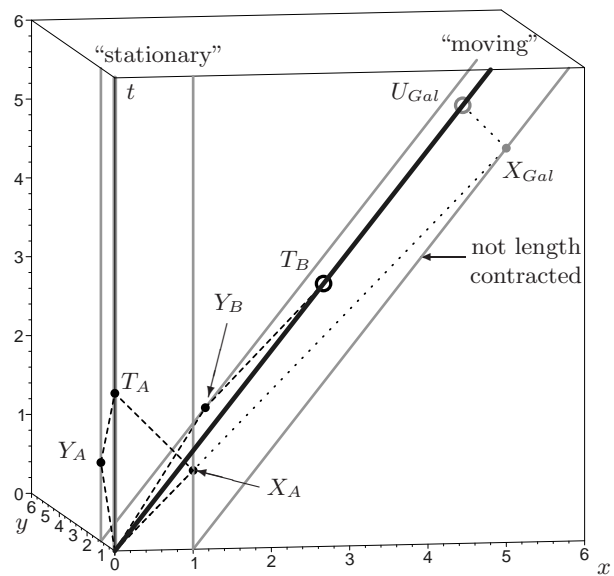}                         
    	}
    \end{picture}
    \hspace*{-207pt}
    \begin{picture}(200,200)(-30,0)
    	\put(53,73){\vector(-4,1){24}}
    \end{picture}
    \vspace*{-0.5in}
\caption{\small
This is A's spacetime diagram of B's
identical apparatus {\em without} length contraction.
Observe that without
length contraction, the light rays reflected by the moving mirrors
are not received simultaneously (at $T_B$) by the moving source.
}
\label{fig:nocontract}
\end{center}
\end{figure}

In passing, we observe that the
length-contraction~factor $(1-(v/c)^2)^{1/2}$,
not found in
the Galilean transformations,%
\cite{fn:GalileanTransformations}
enforces the requirement that the reflection occur
at event~$X_B$ so that the reflected light ray is received
at event~$T_B$.
Without length contraction,
the reflection 
occurs at
event~$X_{Gal}$, and the reception
event~$U_{Gal}$
occurs 
at the source {\em after} event~$T_B$.
(See Figure~\ref{fig:nocontract}.)
Such a result 
violates the principle of relativity since
one's inertial state of motion could now be detected.
Indeed, 
the Michelson-Morley apparatus was used
to measure the time difference between
events~$U_{Gal}$ and $T_B$, as
predicted by the Galilean transformations.
However, no time difference was experimentally observed.%
\cite{Michelson, Panofsky}

}

{
\samepage
For clarity, it is useful to introduce the standard abbreviations.
Let $\gamma$ denote the time-dilation factor
\begin{eqnarray}
    \gamma&=& \frac{1}{\sqrt{1-(v/c)^2}},\label{eq:gamma}
\end{eqnarray}
and let $k$ denote the Doppler-Bondi
factor%
\cite{Ellis, Bondi} %
\begin{eqnarray}
    k&=& \sqrt{\frac{1+(v/c)}{1-(v/c)}}.\label{eq:k}
\end{eqnarray}
With these abbreviations, the coordinates
of $X_B$, $Y_B$, and $T_B$ can be expressed as
\begin{eqnarray*}
X_B:&\quad& \left( kL  ,
\quad 0, \quad k\frac{L}{c} \right)\\
Y_B:&\quad& \left(  \gamma v \frac{L}{c} ,
 \quad L ,\quad
 \gamma\frac{L}{c} \right)\\
T_B:&\quad& \left(  2\gamma v \frac{L}{c} ,
\quad 0,\quad 2\gamma\frac{L}{c} \right).
\end{eqnarray*}
For the examples used throughout
this paper, the B-frame moves with velocity
$v=0.8c$ relative to the A-frame.  For this choice,
we have
$\gamma=5/3$ and $k=3$.
}

\section{Circular Light Clocks}

\subsection{A generalized apparatus}

Generalizing the analysis of the last section,
it is easy to see that:
\begin{quotation}
\noindent
With {\em any} relative orientation of the arms one
would obtain the same results:
\begin{enumerate}
\item Light rays emitted by the source at event~$O$ to mirrors a
distance $L$
away would be received back at the source
at a time $2L/c$ later.
\item The reflection events at the mirrors are
simultaneous according to that source.
\end{enumerate}
\end{quotation}

So, instead of a {\em pair} of equidistant mirrors, consider
a whole collection of mirrors placed inside a circle
[generally, a sphere]
of radius $L$.
Henceforth, this will be called the ``circular light clock.''%
\cite{Fokker}
What would the spacetime diagram of this light clock look like?

{
\samepage
In this case, one would have a hollow world{\em tube}
to describe the collection
of mirrors for each clock. In addition, for each tick of a given clock,
one would draw the portion of its light cone contained inside
the clock's worldtube.  These cones represent events in spacetime
traced out by the collection of light rays that reflect off
the mirrors from tick to tick. (See Figure~\ref{fig:MM}.)

                      %
\begin{figure}[!b]
\begin{center}
		\resizebox{!}{3.75in}{
\begin{picture}(200,300)(35,0)
	\put(0,0){
		\resizebox{!}{4in}{
		     \includegraphics*[bb=184 37 460 345]{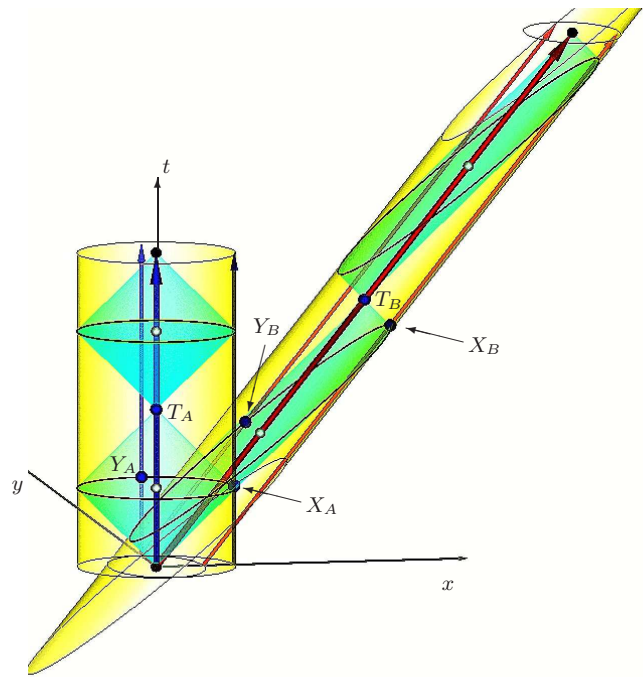}   
		}
	}
\end{picture}
\hspace*{-207pt}
\begin{picture}(200,200)(35,0)
	\put(180,45){$x$}\put(61,188){\vector(0,1){30}}\put(63,220){$t$}\put(0,87){$y$}

	\put(65,117){$T_A$} 				
	
	\put(41,95){$Y_A$} 					

	\put(152,164){$T_B$}  				

		
		\put(101,153){$Y_B$}             
		\put(106,148){\vector(-1,-4){7}} 

		
		\put(123,78){$X_A$}		          
		\put(121,82){\vector(-4,1){24}}   
		                                
		\put(191,145){$X_B$}              
		\put(189,149){\vector(-4,1){24}}  

\end{picture}
}
\caption{\small
Two circular light clocks in relative motion.
For each light clock, the intersection of the
worldtube and the
light cones from two consecutive ticks
is a circle of simultaneous
events for that clock.
The white dots represent events at the source that are
simultaneous with the corresponding circle of intersection.
Note that the ``moving'' circular mirror
is length-contracted in this inertial frame.
}
\label{fig:MM}
\end{center}
\end{figure}
}

In particular,
the stationary light clock will be drawn with
a circular cross-section.  The moving one
will be drawn tilted with an elliptical cross-section
since it is length-contracted in the direction of relative motion.
Then,
given a starting emission
event on the axis of each worldtube,
one traces out the paths of the light rays and their reflections,
which are drawn upward with a slope of~$45$~degrees 
in this diagram.
For simplicity, the starting emission event~$O$
is taken to be
the intersection of the two sources.
(Refer again to Figure~\ref{fig:MM}.)

We now make a series of observations.

The original pair of reflection events $X_A$ and $Y_A$
(and $X_B$ and $Y_B$, respectively)
are among the events on the circle
of mutual intersections of the worldtube
of the light clock
and the light cones of its zeroth and first ticks.
We extend our definition of simultaneity according to
this clock to that circle of events. 
In fact, one can extend this simultaneity
to events on the unique [hyper]plane that contains
this circle [respectively, sphere].  Physically,
this [hyper]plane represents
``all of space at a particular instant for this light clock.''
As before, the events that are simultaneous to this light clock
are generally different from those events 
determined to be simultaneous by the other light clock.

With this notion of simultaneity, the light cones
can be interpreted in a complementary way.
For each light clock,
a simultaneous ``slice'' of
its cones represents a circular [respectively, spherical] wavefront traveling
at the speed of light.
Hence, the light cone can also be interpreted as a sequence of wavefronts
traveling at the speed of light.%
\cite{wavefronts}


\subsection{Visualizing proper-time}

By continuing this light-clock construction along each
inertial observer's worldtube,
an accurate visual representation of the proper
time elapsed for each observer is obtained. 
From such a diagram, however, it may not be evident that the two
observers are equivalent.

\enlargethispage*{25\baselineskip}

{\samepage

First, we demonstrate their
symmetry with the Doppler Effect.
(See Figure~\ref{fig:sym}.)
Suppose these inertial observers emit light signals at one-tick intervals.
From the diagram,
each observer receives those signals from the other observer
at three-tick intervals.  In other words,
the received frequency is one-third of the original frequency,
in accordance with the Doppler effect
for two observers separating with speed $v=0.8c$.
In general,
a light ray emitted by a source
at its first tick after separation
reaches the receiver at the receiver's
$k$th tick
after separation, where $k$
is the Doppler-Bondi factor defined in equation~\ref{eq:k}.
This is the basis of the Bondi $k$-calculus.%
\cite{Ellis, Bondi, Mermin, Marder}

                      %
\begin{figure}[H]
\begin{picture}(200,230)(20,0)
	\put(0,0){
		\resizebox{!}{3in}{
			\includegraphics*[bb=125 40 500 390]{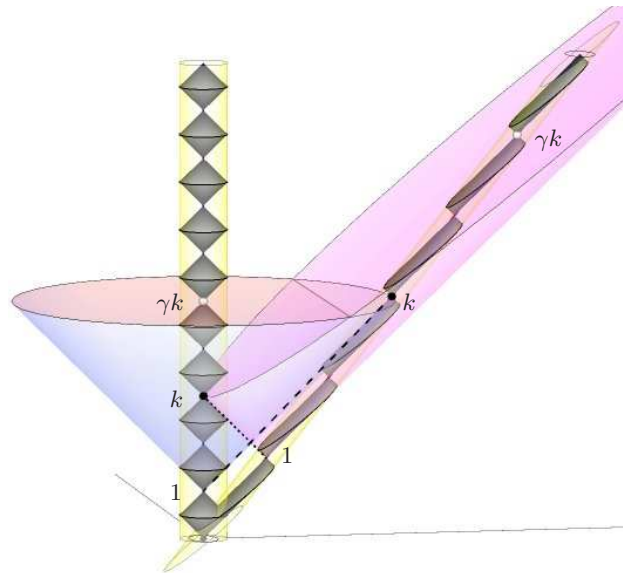}      
		}
	}
\end{picture}
\hspace*{-207pt}
\begin{picture}(200,230)(20,0)
	\put(65,30){$1$}
	\put(65,65){$k$}
	\put(60,102){$\gamma k$}
	\put(107,44){$1$}
	\put(153,102){$k$}
	\put(203,164){$\gamma k$}
\end{picture}
\caption{\small
Symmetry of the observers: The Doppler and Time-Dilation effects.
For a relative speed of $v=0.8c$, we have $\gamma=5/3$ and $k=3$.
}
\label{fig:sym}
\end{figure}
}

Next, we demonstrate their symmetry with the
time-dilation effect.
(Refer again to Figure~\ref{fig:sym}.)
Consider the signal emitted at the first tick.
As just noted, this signal is received by the receiver
at his third tick.  According to the source of that signal,
that distant reception event is simultaneous with his
fifth tick. In other words,
the apparent elapsed time assigned to a distant event
is five-thirds as long as the proper elapsed time
measured by the inertial observer who visits that distant event.
In general, the apparent elapsed time assigned to a distant event is
$\gamma$ times as long as
the proper elapsed time
measured by the inertial observer who meets that distant event,
where $\gamma$ is the time-dilation factor defined in equation~\ref{eq:gamma}.

\begin{figure}[!ht]
\begin{center}
\begin{picture}(200,340)
        \put(0,0){
            \resizebox{!}{2.75in}{
	    \includegraphics*[90pt,35pt][290pt,332pt]{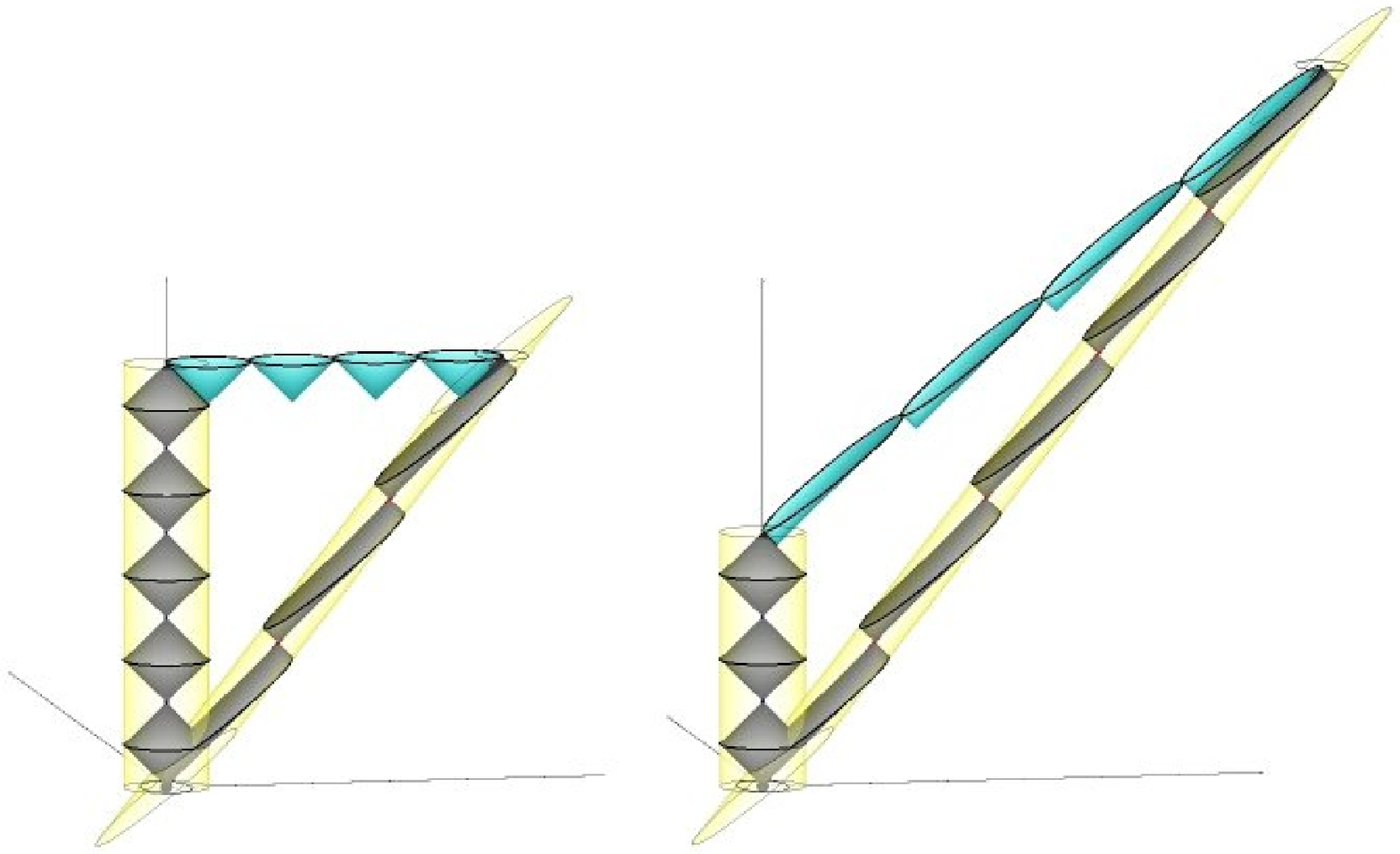}}
	    }
	\put(0,140){
            \resizebox{!}{2.75in}{
	    \includegraphics*[295pt,35pt][540pt,332pt]{SalgadoFig09.eps}}
	    }

\end{picture}
\hspace*{-207pt}
\begin{picture}(200,340)
    \put(0,0)
    {
    \begin{picture}(100,100)(0,0)
	\put(0,40){\begin{turn}{90} apparent-time \end{turn}}
	\put(22,122){apparent-distance}
	\put(50,43){\begin{turn}{51} proper-time \end{turn}}
    \end{picture}
    }
    \put(0,140)
    {
    \begin{picture}(100,100)(0,0)
	\put(72,72){\begin{turn}{51} apparent-time \end{turn}}
	\put(53,117){\begin{turn}{39} apparent-distance \end{turn}}
	\put(0,25){\begin{turn}{90} proper-time \end{turn}}
    \end{picture}
    }
\end{picture}
\caption{\small
Symmetry of the observers: The spacetime analogue of the Pythagorean Theorem:
$(\mbox{proper-time})^2=(\mbox{apparent-time})^2-(\mbox{apparent-distance}/c)^2$.
}
\label{fig:pyth}
\end{center}
\end{figure}

In addition, the source measures the apparent distance
to that reception event to be
$vt_{apparent}=v\cdot(\gamma k\rm\ ticks)$.
For $v=0.8c$, this is
\begin{eqnarray*}
	v\gamma k\rm\ ticks
	=(0.8c)\left(\frac{5}{3}\right)(3)(1\rm\ tick)
	&=&4{\rm\ ticks}\cdot c\\
	&=&
	4\mbox{\ ``light-ticks''}.
\end{eqnarray*}
Since $L$ is the radius of the worldtube and
$1 {\rm\ tick} = (2L/c)$, this distance can be expressed
as ``four worldtube {\em diameters}.''  
This suggests the diagrams in Figure~\ref{fig:pyth}.
Of course, this is just
the calculation of the square-interval in terms of the temporal and spatial coordinates
\begin{eqnarray*}
	\left(
	     \begin{array}{c}
	     \mbox{proper} \\
	     \mbox{time}
	     \end{array}
	\right)^2 
&=& 
	\left(
	     \begin{array}{c}
	     \mbox{apparent} \\
	     \mbox{time}
	     \end{array}
	\right)^2 
	-
	\frac{1}{c^2}\left(
	     \begin{array}{c}
	     \mbox{apparent} \\
	     \mbox{distance}
	     \end{array}
	\right)^2 
\\
(3{\rm\ ticks})^2&=&(5{\rm\ ticks})^2 - (4{\rm\ ticks})^2 \nonumber,
\label{eq:pythagorean}
\end{eqnarray*}
which can be regarded as the spacetime version of the Pythagorean theorem.
Observe that, for a constant value of the proper-time,
the admissible pairs of temporal and spatial coordinates locate
events on a hyperbola.

                      %

\clearpage

\subsection{The Clock Effect}
\begin{figure}[!hb]
\begin{center}
	\resizebox{3in}{!}{
        	\includegraphics*[bb=110 65 473 743]{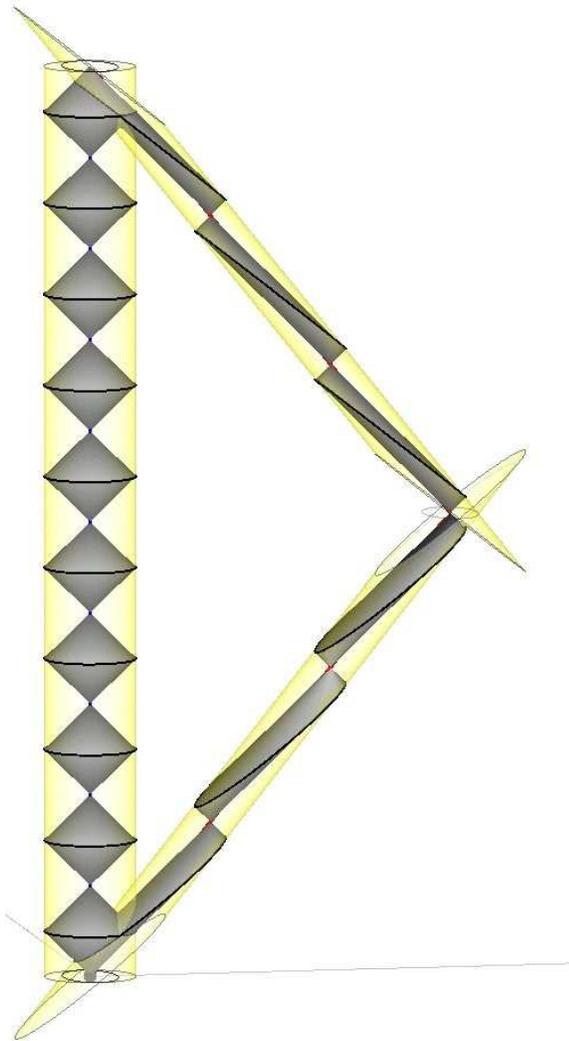}  
	    }
\caption{\small
The Clock Effect.
A non-inertial observer travels away with velocity $v=0.8c$ for 3 ticks,
then returns with velocity $v=-0.8c$ for another 3 ticks.
Between the
departure and reunion events,
he has logged 6 ticks for his entire trip,
whereas the inertial observer has logged 10 ticks.
}
\label{fig:tp}
\end{center}
\end{figure}

With this pictorial device, we present a
visual representation of
the Clock Effect. (See Figure~\ref{fig:tp}.)

                      %

From the diagram,
the non-inertial observer travels away with velocity $v=0.8c$ for 3 ticks,
then returns with velocity $v=-0.8c$ for another 3 ticks.
Between the
departure and reunion events,
he has logged 6 ticks for his entire trip.
On the other hand, the inertial clock logs
\begin{eqnarray*}
	\frac{ ({\rm first\ }3{\rm\ ticks})  }{\sqrt{1-(0.8c/c)^2}}
	+
	\frac{ ({\rm second\ }3{\rm\ ticks}) }{\sqrt{1-(-0.8c/c)^2}}
	=10\rm\ ticks
\end{eqnarray*}
between the same departure and reunion events.
Clearly, the diagram reveals that more time
elapses for the inertial observer that meets both events.

In addition,
the two observers are certainly inequivalent.
The kink in the non-inertial worldline
causes the sequence of simultaneous events to change
discontinuously, leading to the apparent break in the non-inertial worldtube.
This is not to say that the non-inertial worldtube actually breaks.
Rather, it is an artifact of how the diagram was drawn.
In order to draw the true worldtube, a more careful analysis
with a detailed model of the apparatus is needed.
We refer the reader to some articles on the Clock Effect
that discuss this kink in the non-inertial
observer's worldline.%
\cite{Arzelies, Ellis, Marder}%

\subsection{A brief summary}

Let us summarize the logical development up to this point.

Given the simplified Michelson-Morley apparatus in relative motion,
the invariance of the speed of light (so that {\em all} light rays
are drawn at an angle of $45$~degrees) is used to draw the light rays
associated with the perpendicular arm.  Invoking the principle of relativity
(so that the duration of the round-trip
defines the same standard tick), we deduce the effect of time dilation.
Again using the invariance of the speed of light, we draw the light rays
associated with the parallel arm.
Again invoking the principle of relativity,
we deduce the effects of length contraction
and the relativity of simultaneity.

Generalizing these results to arbitrary directions, we obtain the
circular light clock.
By continuing this construction along a piecewise-inertial worldline,
we obtain a visual representation of the proper-time elapsed along
that worldline.


\section{Longitudinal Light Clocks
}

\begin{figure}
\begin{center}
\begin{picture}(200,200)(0,0)
	\put(15,10){$O$}
	\put(13,95){$T_A$}
	\put(65,50){$X_A$}
	\put(140,125){$X_B$}
	\put(115,150){$T_B$}
	\put(45,10){\small $\ell$}
	\put(65,10){\small $L$}
	\resizebox{!}{2.5in}{
		\includegraphics*[2in,3.4in][7in,8.4in]{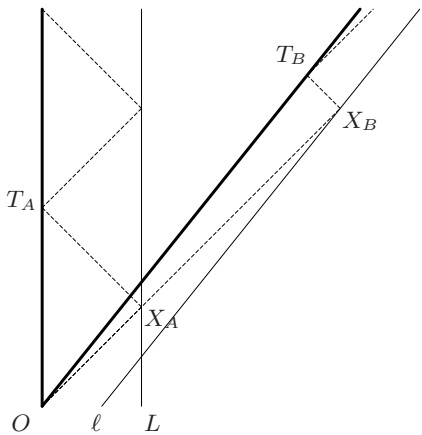}  
	}
\end{picture}
\caption{\small
Two longitudinal light clocks.  The apparent length $\ell$ of the moving
apparatus was shown to be $L(1-(v/c)^2)^{1/2}$.
}
\label{fig:LC}
\end{center}
\end{figure}

\begin{figure}
\begin{center}
\begin{picture}(200,400)(0,0)
	\resizebox{!}{6in}{
		\includegraphics*[1.75in,3in][4.25in,8.4in]{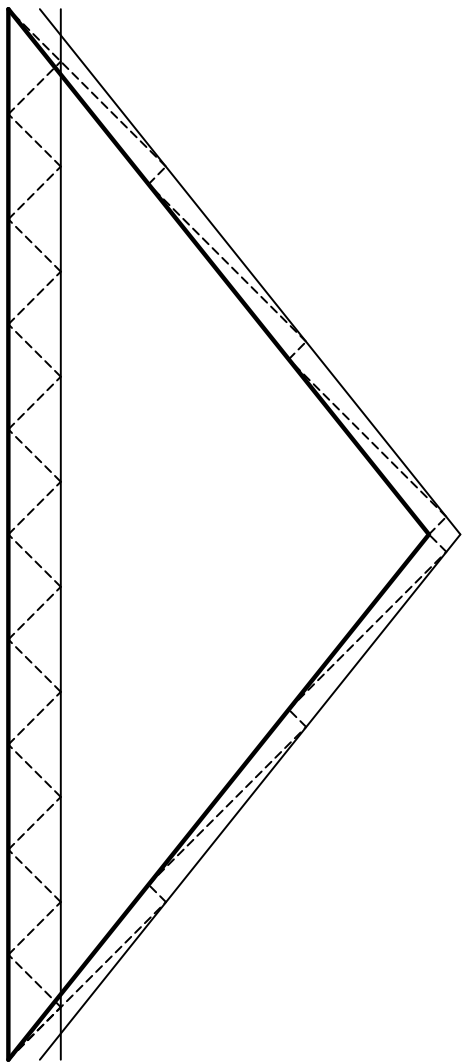}
	}
\end{picture}
\vspace*{-0.25in}
\caption{\small
The Clock Effect with the longitudinal light clocks.
}
\label{fig:twins}
\end{center}
\end{figure}

Let us now consider a simplified
two-dimensional version of the light clock diagram.
Consider the worldlines of the source and of
one longitudinal mirror, that is,
the longitudinal light clock.
As before, the moving light clock
appears length-contracted in the direction of relative motion.
In this case,
we have the following diagrams. (See Figures~\ref{fig:LC} and~\ref{fig:twins}.)

                      %

                      %

Although these figures are much easier to draw, it is unfortunate
that the role of length contraction appears here so prominently.
Recall that event~$T_B$ was determined using 
the invariance of the speed of light and the principle of
relativity, which required that the transverse and longitudinal
reflections be received simultaneously at the source. 
However, without the transverse direction, 
the role of invariance may not
be evident. 

In this section, we will draw attention to 
a certain geometric property of this diagram
and use it to emphasize instead
the ``invariance of the spacetime interval''.

\pagebreak

\subsection{A simple construction}

\begin{figure}[!b]
\begin{center}
\begin{picture}(200,220)(0,0)
	\put(85,85){$O$}
	\put(108,122){$F$}
	\put(95,155){$T_A$}
	\put(145,108){$X_A$}
	\put(159,190){$T_B$}
	\put(180,170){$X_B$}
	\put(80,177){\vector(-1,-2){10}}
	\put(40,180){\shortstack{
		\small hyperbola\hfill
		\\$
		\tau^2=t^2-(x/c)^2$}
	}
	\resizebox{!}{3in}{
		\includegraphics[2in,3.2in][7in,8.4in]{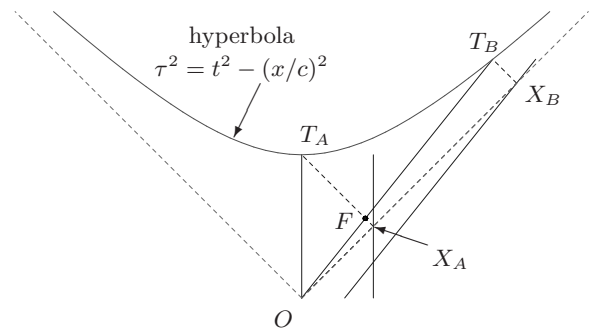}  
	}
\end{picture}
\hspace*{-207pt}
\begin{picture}(200,220)(0,0)
	\put(144,116){\vector(-3,1){20}}
\end{picture}
\vspace*{-1in}
\caption{\small Since $T_A$ and $T_B$ are points of this hyperbola,
the triangles $\triangle OX_AT_A$ and $\triangle OX_BT_B$
have the same area.
Let $k=(OX_B)/(OX_A)$. Using the similarity of
the triangles $\triangle OX_AF$ and $\triangle OX_BT_B$,
it follows that $(X_AF)/(X_AT_A)=1/k^2$.
With these facts,
given $O$, $T_A$, $X_A$, and $F$,
one can easily determine $X_B$ and $T_B$.
(It will be shown that $k$ is precisely
the Doppler-Bondi factor.)
}
\label{fig:hyperbola}
\end{center}
\end{figure}

Refer to the diagram of two longitudinal light clocks.
(See Figure~\ref{fig:hyperbola}.)

                      %

Consider the triangles
$\triangle OX_AT_A$ and $\triangle OX_BT_B$,
which are
formed from the
timelike intervals from event $O$ to the first ticks
and
their associated light rays.
Since, on a spacetime diagram,
events $T_A$ and $T_B$ are at equal intervals from event $O$,
they lie on a rectangular hyperbola
asymptotic to the light cone of event~$O$.
From this, it can be shown that these triangles
(which are related by Lorentz transformations)
have the same area. In fact, we will show in Section~\ref{sec:invariantArea} that
this area is proportional to the square-interval
of one tick.%
\cite{area}%

Using the similarity of triangles $\triangle OX_AF$ and
$\triangle OX_BT_B$ and the relation
$k=(OX_B)/(OX_A)$,  we obtain the useful corollary
that $$\frac{(X_AF)}{(X_AT_A)}=1/k^2.$$
With this property, we can now draw the longitudinal light clock
with the emphasis on the invariance of the interval, rather than
on length contraction.

Given the standard tick for the stationary observer
(events~$O$, $T_A$, and~$X_A$)
and the worldline of a moving observer (line $OF$),
one can determine
the standard tick for the moving observer
(events~$O$, $X_B$ and~$T_B$) as follows.
\begin{itemize}
    \item Measure $(X_AF)/(X_AT_A)$. [For a classroom activity, one might
use a sheet of graph paper with its axes aligned with the future
light cone of event~$O$.]
	\item Calculate $k$ using the corollary $(X_AF)/(X_AT_A)=1/k^2$.
		  (In the next section, we show that $k$ is equal to the Doppler-Bondi factor.)
	\item Determine the reflection event~$X_B$ along the outgoing light ray
			using the relation $k=(OX_B)/(OX_A)$. 
	\item Determine the reception event~$T_B$ 
		  by tracing the reflected light ray back onto the moving worldline.
	       This displays the time dilation effect.
	\item Determine the worldline of the longitudinal mirror 
			by drawing through $X_B$ the parallel to $OF$.
			This displays the length contraction effect.
	\item Finally, determine a set of simultaneous events for this clock  
		by first completing the parallelogram with sides $OX_B$ and $X_B T_B$
		and then drawing the diagonal through $X_B$. 
		This displays the relativity of simultaneity.
\end{itemize}

\subsection{An invariant area}
\label{sec:invariantArea}

The following calculation reveals
that this area of the triangles used in the previous section
is proportional to the square-interval of one tick.%
\cite{area}

Since we will be discussing an aspect of the 
{\em geometry} of the spacetime diagram,
it is convenient to work with a more natural set of coordinates $(x/c, t)$,
where now each coordinate has the same units.

                      %
\begin{figure}[!b]
\begin{center}
    \begin{picture}(200,250)(30,0)
    	\put(212,181){ \makebox(0,0)[l]{$Q=( t, t )$} }
    	\put(20,-7){$O$}
    	\put(165,181){ \makebox(0,0)[r]{$T=( \frac{x}{c} , t )$} }
    	\put(1,14){$S$}
    	\put(196,159){$X$}
    	\put(175,138){$P=( \frac{x}{c} , \frac{x}{c} )$}
    	\put(167,-5){$\frac{x}{c}$}
    	\put(55,135){\vector(1,-2){10}}
    	\put(0,140){\shortstack{
    		\small hyperbola\hfill
    		\\$\tau^2=t^2-(x/c)^2=2\xi\eta$}
    	}
    	\put(-2,0){
    	\resizebox{!}{3in}{
    		\includegraphics{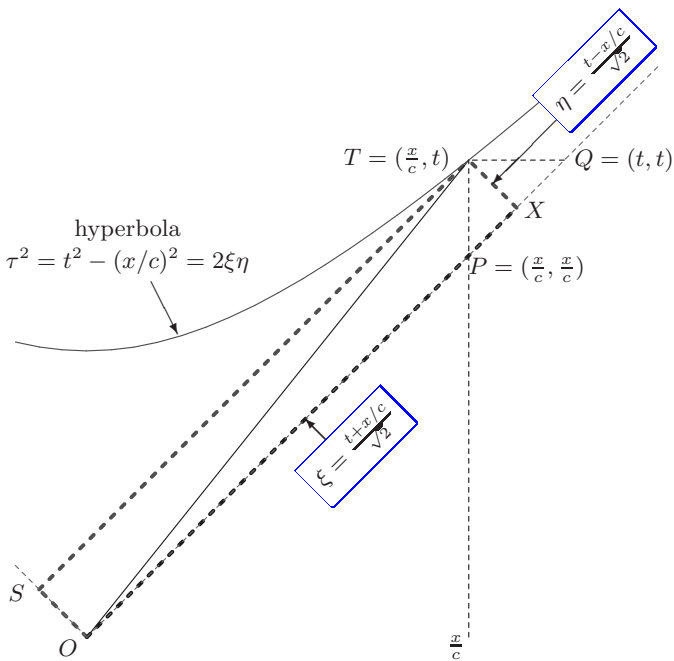}  
    	    }
    	}
    
    \end{picture}
    \hspace*{-207pt}
    \begin{picture}(200,250)(30,0)
    	\put(210,197){\vector(-1,-1){25}}
    	\put(200,197){%
    		\begin{turn}{45}\fcolorbox{blue}{white}{$\eta= \frac{t-x/c}{\sqrt{2}} $} \end{turn}%
    		} 
    	\put(125,72){\vector(-1,1){11}}
    	\put(110,55){%
    		\begin{turn}{45}\fcolorbox{blue}{white}{$\xi= \frac{t+x/c}{\sqrt{2}} $} 
    		
    		\end{turn}%
    		} 
    \end{picture}
\caption{\small The legs of triangle $\triangle OXT$ have
measure\\
$\xi=(2)^{-1/2}(t+x/c)$ and $\eta=(2)^{-1/2}(t-x/c)$.
}
\label{fig:DiracCoords}
\end{center}
\end{figure}

Consider the segment drawn from the emission event~$O$ to 
any event $T$ with coordinates $(x/c,t)$.
(See Figure~\ref{fig:DiracCoords}.)
Regard that segment as the hypotenuse of a Euclidean right triangle whose sides
are parallel to the light cone of event~$O$. The legs of this triangle have
measure
$$\xi=\frac{t+x/c}{\sqrt{2}}\qquad \eta=\frac{t-x/c}{\sqrt{2}}.$$
These are called the Dirac light-cone coordinates%
\cite{area, DiracCoords}
of event~$T$ in the A-frame.
In these coordinates, the Euclidean area of this triangle is simply $\xi\eta/2$,
which is equal to $(t^2-(x/c)^2)/4$.
That is, the Euclidean area of triangle $\triangle OXT$
is equal to one-fourth of the square-interval from event~$O$ to event~$T$.

Let us explicitly verify that this area is invariant under Lorentz transformations.
\cite{fn:unitDeterminant}
Using Equation~\ref{eq:LT},
the light-cone coordinates of event~$T$ in the B-frame are
\begin{eqnarray*}
    \xi'= \frac{t'+x'/c}{\sqrt{2}}
     & =& \gamma\left(1-\frac{v}{c}\right)
    \left(
            \frac{t+x/c}{\sqrt{2}}
    \right)\\
    & =& k^{-1} \left(\frac{t+x/c}{\sqrt{2}}\right)\ =\ k^{-1}\xi\\
    &&\\
    \eta' = \frac{t'-x'/c}{\sqrt{2}}
    &=& \gamma\left(1+\frac{v}{c}\right)
    \left(
            \frac{t-x/c}{\sqrt{2}}
    \right)\\
    & =& k \left( \frac{t-x/c}{\sqrt{2}} \right)\ =\ k \eta,
\end{eqnarray*}
where $k$ is Doppler-Bondi factor defined in Equation~\ref{eq:k}.
Thus, the quantity $\xi\eta/2$ is Lorentz invariant.

It is instructive to interpret this geometrically.
For concreteness, let us start with $T=T_B$ and $X=X_B$. 
In this case, we seek the Lorentz transformation that
sets $OT_B$ to be at rest in the B-frame. Imagine sliding $T_B$ down along the
hyperbola until $OT_B$ is vertical. As that happens, $X_B$ slides down the light cone,
scaling the $\xi$-leg down by a factor~$k$ and 
scaling the $\eta$-leg up by a factor~$k$. 
Thus, the area of triangle $\triangle OX_BT_B$ is invariant.

We extend this result to the parallelogram formed with light rays $OX_B$ and $X_BT_B$,
{\em i.e.,} the {\em``area of intersection between 
the interior of $O$'s future light-cone and 
the interior of $T_B$'s past light-cone.''} 
Clearly, this area is invariant and is proportional to the 
square-interval of one tick. 

In fact, this result generalizes to higher-dimensions:
the analogous {\em volume} of intersection between the light-cone interiors
is also invariant and is proportional to the square-interval of one tick.
We show this for the 3- and 4-dimensional case.
The Euclidean volume of a cone with an
elliptical base is $(\pi ab)h/3$,
where $a$ and $b$ are the semi-major and semi-minor axes of the elliptical base, 
and $h$ is the altitude of the cone. 
In four dimensions, the hypervolume of a cone
with an ellipsoidal base is $(4\pi ab^2/3)h/4$.
Since we orient the base so that 
$b$ is a length along the transverse direction,
it is unchanged under a Lorentz transformation.
Regarding $\triangle OSX$ in Figure~\ref{fig:DiracCoords} 
as the $xt$-cross-section of the light cone of event~$O$, 
observe that its area is $(2a)h/2=ah$.
Using the symmetries of the parallelogram, 
the area of $\triangle OSX$ is equal to 
the area of $\triangle OXT$.
Thus, $ah$ is invariant. It follows that the
volume of intersection between the light-cone interiors
is invariant.

\section{Final Remarks}

By drawing the spacetime diagram of a Michelson-Morley apparatus,
we have obtained an accurate visualization of the proper-time elapsed
along a piecewise-inertial observer's worldline.
Measurements of spacetime intervals have been reduced to the counting of ticks,
emphasizing the relativistic view that ``time is measured by an observer's clock.''
We believe that the resulting diagram can be used to discuss special relativity
in a qualitative way which emphasizes the physics first and the algebra second.
We feel this could easily be incorporated into the standard textbook treatments 
of special relativity, which often discuss the Michelson-Morley experiment.

The ideas presented in this paper are being implemented
in a series of interactive computer programs
which will be posted to our website.
\cite{fn:url}

\section{Acknowledgments}

Part of this work was supported by a Truman State University
Curriculum Development Grant.


\begin{thebibliography}{99}
\bibitem{fn:ignoreMomentum}
	Implicitly, it is assumed that the impulses transferred to the mirrors
	upon reflection are negligible so that
	these mirrors maintain their separation from the source.
\bibitem{fn:noTransverseContraction}
	Strictly speaking, we should prove that, for a moving object, lengths along
	its transverse directions are unchanged.
	We refer the reader to Feynman
	\cite{FeynmanLectures},
	Mermin
	\cite{MerminSTSR}, and
	Arons
	\cite{Arons}
	for a symmetry argument.
\bibitem{fn:unitDeterminant}
	Indeed, since a proper Lorentz transformation has unit determinant, 
	spacetime volume is invariant.%
\end{thebibliography}
\end{document}